\documentclass[12pt]{article}
\usepackage{amsmath}

\usepackage{caption2}
\begin{document}

\title{\vspace{-.60cm}\textbf{\Large Local hidden-variable models and \\
\vspace{-.2cm} negative-probability measures}}
\vspace{1cm}
\author{Jos\'{e} L. Cereceda  \\ 
\textit{C/Alto del Le\'{o}n 8, 4A, 28038 Madrid, Spain}  \\
\small{Electronic address: jl.cereceda@teleline.es}}

\date{December 13, 2000}

\maketitle
\begin{abstract}
Elaborating on a previous work by Han et al., we give a general, basis-independent proof of the necessity of negative probability measures in order for a class of local hidden-variable (LHV) models to violate the Bell-CHSH inequality. Moreover, we obtain general solutions for LHV-induced probability measures that reproduce any consistent set of probabilities.

\vspace{.3cm}

\end{abstract}
\vspace{.75cm}

In 1982, M\"{u}ckenheim [1] made use of negative probability functions in an attempt to resolve the Einstein-Podolsky-Rosen (EPR) paradox [2]. For this purpose, M\"{u}ckenheim built a classical model endowed with negative probabilities that reproduced all the statistical predictions of quantum theory for the singlet state of two spin-half particles. While the physical meaning of extended probabilities is far from obvious [3], this attempted solution of the EPR paradox might seem [4], ``$\ldots$as unattractive as ({\em but not more unattractive than}) all the others'' (present author's emphasis). Subsequently, Home, Lepore and Selleri [5] put forward a general argument demonstrating that one can always reproduce the quantum mechanical results for nonfactorizable state vectors of correlated systems by means of probabilities of the Clauser-Horne type [6] provided one allows for probabilities not obeying Kolmogorov's axiom according to which probabilities $p$ are restricted to the range $0\leq p \leq 1$. More recently, Han, Hwang and Koh [7] obtained explicit solutions for classical probability measures that reproduce quantum mechanical predictions for some spin-measurement directions for all entangled states, and proved the necessity of negative probability measures in this case. In the present paper, we shall extend the proof by Han et al.\ in the following sense. While Han et al.'s proof relies in a special basis (see Eqs.\ (28) and (39) of [7]) to show the negativity of probability measures for the considered local hidden-variable (LHV) model, ours proves the necessity of such negative probability measures in {\em all\/} instances where the predictions of the LHV model are made to violate the Bell-CHSH inequality [8-10]. This is done without relying on any particular basis states or measurement directions. In fact, our result holds irrespective of whatever quantum mechanical consideration. Moreover, we give general solutions for LHV probability measures that reproduce {\em any\/} conceivable set of probabilities satisfying certain requirement conditions, namely, the normalization condition and the causal communication constraint (cf.\ Eqs.\ (17) and (28)-(29) below). A set of probabilities fulfilling these requirement conditions will be referred to as a consistent set. The proof goes in a rather straightforward way as follows.

Consider an experiment of the EPR type designed to test the Bell-CHSH inequality. Two correlated particles 1 and 2 fly apart in opposite directions from some common source. Subsequently, each of the particles enters its own measuring apparatus which can measure either one of two physical variables at a time---$a_1$ or $a_2$ for particle 1 and $b_1$ or $b_2$ for particle 2. The possible values of these variables may be taken to be $+1$ and $-1$. The source emits a very large number of particle pairs. The basic entity to be considered is the joint probability $p(a_j =m,b_k =n)$ that the outcome of the measurement of $a_j$ on particle 1 is $m$, and that the outcome of the measurement of $b_k$ on the paired particle 2 is $n$, where $j,k = 1,2$, and $m,n=\pm 1$. A representative deterministic LHV model describing this experiment is as follows [7,11]. The main assumption made by such a model is that, for every pair of particles emitted by the source, there exists a hidden variable $\lambda$ (with domain of variation $\Lambda$) which determines locally (for example, at the common source) the response of the particles to each of the measurements they can be subjected to. For the experiment under consideration, the set of all $\lambda$ can then be partitioned into 16 disjoint subsets $\Lambda_i$ (with respective probability measure $m_i$) according to the outcomes of the four possible measurements, $a_1$ and $a_2$ for particle 1 and $b_1$ and $b_2$ for particle 2. In Table 1 we display the 16 rows characterizing the subsets $\Lambda_i$. The {\em i\/}th row indicates the response of the particles to the different measurements when the particle pair is described by a hidden variable pertaining to the subset $\Lambda_i$. So, for example, if a particle pair is described by a given $\lambda \in \Lambda_2$, then the particles must behave according to the following local plan: if $a_1$ is measured on particle 1 the result will be $+1$, if $a_2$ is measured on particle 1 the result will be $+1$, if $b_1$ is measured on particle 2 the result will be $+1$, and if $b_2$ is measured on particle 2 the result will be $-1$. (Note that, for each of the plans, the agreed result for $a_j$ is independent of which measurement ($b_1$ or $b_2$) is performed on particle 2, and similarly the agreed result for $b_k$ is independent of which measurement ($a_1$ or $a_2$) is performed on particle 1.)

\begin{table}[t]
\caption{\footnotesize The 16 possible subsets into which the total set $\Lambda$ can be partitioned. The hidden variables in each subset $\Lambda_i$ uniquely determine the outcomes for each of the four possible single measurements $a_1$, $b_1$, $a_2$, and $b_2$.}
\vspace{2mm}
\begin{center}
\begin{tabular}{c|cccc|c}\hline
\small{Subset of $\Lambda$} & $a_1$ & $b_1$ & $a_2$ & $b_2$ 
& \small{Probability measure}  \\  \hline
$\Lambda_1$ & $+$ & $+$ & $+$ & $+$ & $m_1$  \\  
$\Lambda_2$ & $+$ & $+$ & $+$ & $-$ & $m_2$  \\  
$\Lambda_3$ & $+$ & $+$ & $-$ & $+$ & $m_3$  \\  
$\Lambda_4$ & $+$ & $+$ & $-$ & $-$ & $m_4$  \\  
$\Lambda_5$ & $+$ & $-$ & $+$ & $+$ & $m_5$  \\  
$\Lambda_6$ & $+$ & $-$ & $+$ & $-$ & $m_6$  \\  
$\Lambda_7$ & $+$ & $-$ & $-$ & $+$ & $m_7$  \\  
$\Lambda_8$ & $+$ & $-$ & $-$ & $-$ & $m_8$  \\  
$\Lambda_9$ & $-$ & $+$ & $+$ & $+$ & $m_9$  \\  
$\Lambda_{10}$ & $-$ & $+$ & $+$ & $-$ & $m_{10}$  \\  
$\Lambda_{11}$ & $-$ & $+$ & $-$ & $+$ & $m_{11}$  \\  
$\Lambda_{12}$ & $-$ & $+$ & $-$ & $-$ & $m_{12}$  \\  
$\Lambda_{13}$ & $-$ & $-$ & $+$ & $+$ & $m_{13}$  \\  
$\Lambda_{14}$ & $-$ & $-$ & $+$ & $-$ & $m_{14}$  \\  
$\Lambda_{15}$ & $-$ & $-$ & $-$ & $+$ & $m_{15}$  \\  
$\Lambda_{16}$ & $-$ & $-$ & $-$ & $-$ & $m_{16}$  \\  \hline
\end{tabular}
\end{center}
\end{table}

From Table 1, we can readily compute the predictions that our LHV model makes for the various probabilities $p(a_j =m,b_k =n)$. These are given by
\begin{align}
p_1 &\equiv p(a_1+;b_1+)=m_1+m_2+m_3+m_4,    \\[-.5mm]
p_2 &\equiv p(a_1+;b_1-)=m_5+m_6+m_7+m_8,    \\[-.5mm]
p_3 &\equiv p(a_1-;b_1+)=m_9+m_{10}+m_{11}+m_{12},     \\[-.5mm]
p_4 &\equiv p(a_1-;b_1-)=m_{13}+m_{14}+m_{15}+m_{16},    \\[-.5mm]
p_5 &\equiv p(a_1+;b_2+)=m_1+m_3+m_5+m_7,    \\[-.5mm]
p_6 &\equiv p(a_1+;b_2-)=m_2+m_4+m_6+m_8,    \\[-.5mm]
p_7 &\equiv p(a_1-;b_2+)=m_9+m_{11}+m_{13}+m_{15},     \\[-.5mm]
p_8 &\equiv p(a_1-;b_2-)=m_{10}+m_{12}+m_{14}+m_{16},     \\[-.5mm]
p_9 &\equiv p(a_2+;b_1+)=m_1+m_2+m_9+m_{10},    \\[-.5mm]
p_{10} &\equiv p(a_2+;b_1-)=m_5+m_6+m_{13}+m_{14},    \\[-.5mm]
p_{11} &\equiv p(a_2-;b_1+)=m_3+m_4+m_{11}+m_{12},     \\[-.5mm]
p_{12} &\equiv p(a_2-;b_1-)=m_7+m_8+m_{15}+m_{16},   \\[-.5mm]
p_{13} &\equiv p(a_2+;b_2+)=m_1+m_5+m_9+m_{13},    \\[-.5mm]
p_{14} &\equiv p(a_2+;b_2-)=m_2+m_6+m_{10}+m_{14},    \\[-.5mm]
p_{15} &\equiv p(a_2-;b_2+)=m_3+m_7+m_{11}+m_{15},     \\[-.5mm]
p_{16} &\equiv p(a_2-;b_2-)=m_4+m_8+m_{12}+m_{16},  
\end{align}
in obvious notation. We are assuming throughout this paper ideal behavior of the measuring apparata and, in particular, perfect efficiency of the detection equipment. This requires the probabilities $p(a_j =m,b_k =n)$ to satisfy the normalization condition
\begin{equation}
\sum_{m,n =\pm 1}  p(a_j =m, b_k =n) =1 ,
\end{equation}
for any $j,k = 1,2$. From Eqs.\ (1)-(4), this in turn implies
\begin{equation}
\sum_{i=1}^{16} m_i =1.
\end{equation}
For convenience for what follows we define the following two quantities,
\begin{equation}
\Sigma_1 = m_4+m_5+m_6+m_8+m_9+m_{11}+m_{12}+m_{13} \, , 
\end{equation}
and
\begin{equation}
\Sigma_2 = 1-\Sigma_1 = m_1+m_2+m_3+m_7+m_{10}+m_{14}+m_{15}+m_{16}\, .  
\end{equation}

Let us now consider the sum of correlations
\begin{equation}
\Delta = c(a_1,b_1) + c(a_1,b_2) + c(a_2,b_1) - c(a_2,b_2),
\end{equation}
entering into the Bell-CHSH inequality [8-10], $|\Delta|\leq 2$, with the correlation coefficient $c(a_j,b_k)$ being given by
\begin{align}
c(a_j ,b_k) = \;\, &p(a_j =1,b_k=1) + p(a_j=-1,b_k=-1)  \nonumber  \\
&-p(a_j=1,b_k=-1) - p(a_j=-1,b_k=1)\,. 
\end{align}
Substituting this in Eq.\ (21), and taking into account the normalization condition in Eq.\ (17), the quantity $\Delta$ can equivalently be written in the form
\begin{equation}
\Delta = 2(p_1+p_4+p_5+p_8+p_9+p_{12}+p_{14}+p_{15}-2),
\end{equation}
where we have used the abbreviated notation introduced in Eqs.\ (1)-(16). Now, by replacing the probabilities appearing in Eq.\ (23) by their respective expressions in Eqs.\ (1)-(16), one obtains the prediction that our LHV model makes for the Bell-CHSH sum of correlations,
\begin{equation}
\Delta_{\text{LHV}} = 2 (1-2\Sigma_1) = 2 (2\Sigma_2 -1).
\end{equation}

The Bell-CHSH inequality is violated whenever $|\Delta| > 2$. Then, from Eq.\ (24), it follows at once that in order for the LHV model to give a violation of the Bell-CHSH inequality it is necessary that either $\Sigma_1 <0$ or $\Sigma_1 >1$ (or, correspondingly, that either $\Sigma_2 >1$ or $\Sigma_2 <0$). In any case, to have \mbox{$|\Delta_{\text{LHV}}|> 2$}, it is necessary that either $\Sigma_1$ or $\Sigma_2$ be negative. Thus, as the negativity of either $\Sigma_1$ or $\Sigma_2$ implies the negativity of at least one of the $m_i$'s, we have proved the necessity of negative probability measures for the considered LHV model if this is to violate the Bell-CHSH inequality. On the other hand, from Eq.\ (24), it is also clear that the prediction by the LHV model does satisfy the Bell-CHSH inequality if, and only if, the condition \mbox{$0\leq \Sigma_1 \text{,} \,\Sigma_2 \leq 1$} is satisfied. Needless to say, the usual case where $\sum m_i =1$ and $0\leq m_i \leq 1$ fulfils this latter condition, and hence, for such a case, $|\Delta_{\text{LHV}}|\leq 2$. This shows that the role played by the assumption of nonnegativity of the probability measures in deriving the Bell-CHSH inequality is as fundamental as the role played by the assumption of local realism. It should be noticed, however, that, although necessary, the requirement of negative probability measure is {\em not\/} a sufficient condition in order to have $|\Delta_{\text{LHV}}|> 2$. Indeed, it may be the case that some of the $m_i$'s be negative, and yet having that \mbox{$0\leq \Sigma_1 \text{,} \,\Sigma_2 \leq 1$}.

Now we are going to give general solutions for LHV probability measures that reproduce any consistent set of probabilities $\{p_1,p_2,\ldots,p_{16}\}$. To this end, let us first consider the special case where $\Delta =2\sqrt{2}$. This value for $\Delta$ can be reproduced by the LHV model if $\Sigma_1 =(1-\sqrt{2})/2$ and $\Sigma_2 =(1+\sqrt{2})/2$. An immediate, ``equally-distributed'' solution of these equations is
\begin{equation}
m_4=m_5=m_6=m_8=m_9=m_{11}=m_{12}=m_{13}=\frac{1-\sqrt{2}}{16},  
\end{equation}
and
\begin{equation}
m_1=m_2=m_3=m_7=m_{10}=m_{14}=m_{15}=m_{16}=\frac{1+\sqrt{2}}{16}.  
\end{equation}
Substituting these values in Eqs.\ (1)-(16) yields the following {\em positive\/} probabilities predicted by the LHV model for the special case considered, $p_1=p_4=p_5=p_8=p_9=p_{12}=p_{14}=p_{15}=(2+\sqrt{2})/8$ and $p_2=p_3=p_6=p_7=p_{10}=p_{11}=p_{13}=p_{16}=(2-\sqrt{2})/8$. It is to be noted that these values for $p_1,p_2,\ldots,p_{16}$ are the same as those predicted by quantum mechanics (QM) in the case that the quantum Bell-CHSH sum of correlations attains the  Cirel'son limit   $\Delta_{\text{QM}}=2\sqrt{2}$ [12]. It should be added that, however, the equally-distributed solution, namely that for which $m_4=m_5=m_6=m_8=m_9=m_{11}=m_{12}=m_{13}$ and $m_1=m_2=m_3=m_7=m_{10}=m_{14}=m_{15}=m_{16}$, is clearly too restrictive since, as may readily be checked from Eqs.\ (1)-(16), it invariably leads to the prediction that $p_1=p_4=p_5=p_8=p_9=p_{12}=p_{14}=p_{15}$ and $p_2=p_3=p_6=p_7=p_{10}=p_{11}=p_{13}=p_{16}$. Consequently, except for the case where these conditions on the probabilities are met, the equally-distributed solution cannot account for the generic set of probabilities $\{p_1,p_2,\ldots,p_{16}\}$. Anyway, it is nevertheless important to realize that, if such probabilities $p_1,p_2,\ldots,p_{16}$ are to be given by the LHV predictions on the right-hand side of Eqs.\ (1)-(16), then the probabilities themselves must obey certain requirement conditions. Specifically, if we want the generic probabilities $p_1,p_2,\ldots,p_{16}$ to be cast into the form displayed by Eqs.\ (1)-(16), then the following relationships between them must necessarily hold
\begin{align}
p_2 &= \frac{1}{2} (1-p_1-p_4+p_5-p_8-p_9+p_{12}+p_{14}-p_{15}),  \notag  \\
p_3 &= \frac{1}{2} (1-p_1-p_4-p_5+p_8+p_9-p_{12}-p_{14}+p_{15}),  \notag  \\
p_6 &= \frac{1}{2} (1+p_1-p_4-p_5-p_8-p_9+p_{12}+p_{14}-p_{15}),  \notag  \\
p_7 &= \frac{1}{2} (1-p_1+p_4-p_5-p_8+p_9-p_{12}-p_{14}+p_{15}),    \\
p_{10} &= \frac{1}{2} (1-p_1+p_4+p_5-p_8-p_9-p_{12}+p_{14}-p_{15}),  \notag  \\
p_{11} &= \frac{1}{2} (1+p_1-p_4-p_5+p_8-p_9-p_{12}-p_{14}+p_{15}),  \notag  \\
p_{13} &= \frac{1}{2} (1-p_1+p_4+p_5-p_8+p_9-p_{12}-p_{14}-p_{15}),  \notag  \\
p_{16} &= \frac{1}{2} (1+p_1-p_4-p_5+p_8-p_9+p_{12}-p_{14}-p_{15}).  \notag  
\end{align}
We note, incidentally, that the set of conditions in Eq.\ (27) is equivalent to the conjunction of the normalization condition in Eq.\ (17) and the so-called causal communication constraint [13]. This latter consistency condition requires that [14]
\begin{align}
& \sum_{n=\pm 1} p(a_j =m, b_1 =n) = \sum_{n=\pm 1} p(a_j =m, b_2 =n),  \\
& \sum_{m=\pm 1} p(a_1 =m, b_k =n) = \sum_{m=\pm 1} p(a_2 =m, b_k =n), 
\end{align}
for any $j,k = 1,2$ and $m,n=\pm1$, and prevents the acausal exchange of classical information between the two parties involved in the EPR experiment. Both quantum mechanics and our LHV model satisfy the requirement of causal communication and hence the predictions by such theories do satisfy each of the constraints in Eq.\ (27).

In searching for a general solution for LHV probability measures $m_1,m_2,\ldots,\linebreak m_{16}$ that reproduces the probabilities $p_1,p_2,\ldots,p_{16}$, what really matters is the fact that, as can be seen from Eq.\ (27), only eight of such probabilities are independent. We may take the independent probabilities to be $p_1$, $p_4$, $p_5$, $p_8$, $p_9$, $p_{12}$, $p_{14}$, and $p_{15}$. Therefore, to invert the set of Eqs.\ (1)-(16), it suffices to consider the eight equations (1), (4), (5), (8), (9), (12), (14), and (15), {\em plus\/} the normalization condition in Eq.\ (18). We are thus left with a system of 9 equations with 16 unknowns $m_1,m_2,\ldots,m_{16}$, in which the probabilities $p_1$, $p_4$, $p_5$, $p_8$, $p_9$, $p_{12}$, $p_{14}$, $p_{15}$ are treated as given parameters. This system determines 9 probability measures as a function of the remaining 7 probability measures and the 8 probabilities above. So, for example, we can get the following general solution for which the set of probability measures $\{m_1,m_4,m_5,m_6,m_8,m_9,m_{11},m_{12},m_{13}\}$ is given in terms of the remaining set $\{m_2,m_3,m_7,m_{10},m_{14},m_{15},m_{16}\}$ and the eight probabilities $p_1$, $p_4$, $p_5$, $p_8$, $p_9$, $p_{12}$, $p_{14}$, and $p_{15}$,
\begin{multline}
\,m_1 = \frac{1}{2} (-1-2m_2-2m_3-2m_7-2m_{10}-2m_{14}-2m_{15}-2m_{16}   \\[-.1cm]
+p_1+p_4+p_5+p_8+p_9+p_{12}+p_{14}+p_{15} ), \quad  
\end{multline}\\[-1.7cm]
\begin{multline}
\,m_4 = \frac{1}{2} ( 1+2m_7+2m_{10}+2m_{14}+2m_{15}+2m_{16}+p_1-p_4   \\[-.1cm]
-p_5-p_8-p_9-p_{12}-p_{14}-p_{15} ), \quad
\end{multline}\\[-1.7cm]
\begin{multline}
\,m_5 = \frac{1}{2} ( 1+2m_2+2m_{10}+2m_{14}+2m_{15}+2m_{16}-p_1-p_4   \\[-.1cm]
+p_5-p_8-p_9-p_{12}-p_{14}-p_{15} ),  \quad
\end{multline}\\[-1.6cm]
\begin{equation}
m_6 = -m_2-m_{10}-m_{14}+p_{14},  
\qquad\qquad\qquad\qquad\quad\quad\quad\quad\quad\quad\quad\quad\quad\,\,\,\,\,\,\,\,\,\,\,\,
\end{equation}\\[-1.6cm]
\begin{equation}
m_8 = -m_7-m_{15}-m_{16}+p_{12},  
\qquad\qquad\qquad\qquad\quad\quad\quad\quad\quad\quad\quad\quad\quad\,\,\,\,\,\,\,\,\,\,\,\,
\end{equation}\\[-1.6cm]
\begin{multline}
\,m_9 = \frac{1}{2} ( 1+2m_3+2m_7+2m_{14}+2m_{15}+2m_{16}-p_1-p_4   \\[-.1cm]
-p_5-p_8+p_9-p_{12}-p_{14}-p_{15} ),  \quad
\end{multline}\\[-1.6cm]
\begin{equation}
m_{11} = -m_3-m_7-m_{15}+p_{15},  
\qquad\qquad\qquad\quad\quad\qquad\qquad\quad\quad\quad\quad\quad\,\,\,\,\,\,\,\,\,\,\,\,\,\,\,
\end{equation}\\[-1.6cm]
\begin{equation}
m_{12} = -m_{10}-m_{14}-m_{16}+p_8,  
\qquad\qquad\qquad\qquad\qquad\quad\quad\quad\quad\quad\quad\quad\,\,\,\,\,\,\,\,\,\,\,\,\,
\end{equation}\\[-1.6cm]
\begin{equation}
m_{13} = -m_{14}-m_{15}-m_{16}+p_4.  
\qquad\qquad\qquad\qquad\quad\quad\quad\quad\quad\quad\quad\quad\quad\,\,\,\,\,\,\,\,\,\,\,\,\,\,
\end{equation}
By inserting the $m_i$'s given by Eqs.\ (30)-(38) into Eqs.\ (1)-(16), and recalling the relations in Eq.\ (27), one can reproduce whichever consistent set of probabilities $\{p_1,p_2,\ldots,p_{16}\}$. Indeed, as may easily be verified, the sum $m_1+m_2+m_3+m_4$, with $m_1$ and $m_4$ given by Eqs.\ (30) and (31), respectively, gives $p_1$. Likewise, the sum $m_5+m_6+m_7+m_8$, with $m_5$, $m_6$, and $m_8$ given by Eqs.\ (32), (33), and (34), respectively, gives $p_2 =(1-p_1-p_4+p_5-p_8-p_9+p_{12}+p_{14}-p_{15})/2$, etc. Similarly, from Eqs.\ (31)-(38), we may also check that the sum $m_4+m_5+m_6+m_8+m_9+m_{11}+m_{12}+m_{13}$ amounts to
\begin{equation}
\Sigma_1 = \frac{1}{2} (3-p_1-p_4-p_5-p_8-p_9-p_{12}-p_{14}-p_{15}).
\end{equation}
Hence, from Eqs.\ (24) and (39), we obtain the LHV prediction $\Delta_{\text{LHV}} = 2(p_1+p_4+p_5+p_8+p_9+p_{12}+p_{14}+p_{15}-2)$, thereby reproducing the generic Bell-CHSH sum of correlations in Eq.\ (23). Of course, according to our previous result following Eq.\ (24), a violation of the Bell-CHSH inequality necessarily implies the quantity $\Sigma_1$ in Eq.\ (39) to be either $\Sigma_1 <0$ or $\Sigma_1 >1$. We note the remarkable fact that there remain seven degrees of freedom in solutions for LHV probability measures reproducing any consistent set of probabilities. For the general solution displayed in Eqs.\ (30)-(38), these degrees of freedom correspond to the variables $m_2$, $m_3$, $m_7$, $m_{10}$, $m_{14}$, $m_{15}$, and $m_{16}$.

We conclude by discussing the case where, due to perfect correlation between the particles, two of the probabilities, say $p_2$ and $p_3$, are equal to zero. This means that the results for the joint measurement of the observables $a_1$ and $b_1$ must both be either $+1$ or $-1$. Thus, from a physical point of view, it is reasonable to suppose that, for the case in which $p_2=0$ and $p_3=0$, the probability measures $m_5$, $m_6$, $m_7$, $m_8$, $m_9$, $m_{10}$, $m_{11}$, and $m_{12}$ do equally vanish (see Eqs.\ (2) and (3)). Otherwise, the LHV model could yield joint detection events which, by assumption, never happen. (Of course, mathematically, we may have $p_2=0$ without actually requiring that $m_5=m_6=m_7=m_8=0$. This will happen, for example, whenever $m_5+m_6=-m_7-m_8$.) On the other hand, the fact that $p_2=p_3=0$ imposes further constraints on the probabilities. Specifically, since $p_1+p_2+p_3+p_4=1$, we have that $p_1=1-p_4$ whenever $p_2=p_3=0$. In addition to this, the first equation in (27) tells us that $p_5=p_8+p_9-p_{12}-p_{14}+p_{15}$ whenever $p_1=1-p_4$ and $p_2=0$. Then, putting $m_5=m_6=m_7=m_8=m_9=m_{10}=m_{11}=m_{12}=0$, and substituting $1-p_4$ for $p_1$, and $p_8+p_9-p_{12}-p_{14}+p_{15}$ for $p_5$ in the general equations (30)-(38), we obtain the following solution for the remaining LHV probability measures in the case that perfect correlation develops between the measurement outcomes of the observables $a_1$ and $b_1$,
\begin{align}
m_1 & =-m_{16}+p_8+p_9-p_{14},  \notag  \\
m_2 & =m_{16}-p_8+p_{14},  \notag  \\
m_3 & =m_{16}-p_{12}+p_{15},  \notag  \\
m_4 & =1-m_{16}-p_4-p_9+p_{12}-p_{15},    \\
m_{13} & =m_{16}+p_4-p_8-p_{12},  \notag  \\
m_{14} & =-m_{16}+p_8,       \notag  \\
m_{15} & =-m_{16}+p_{12},  \notag  
\end{align}
where now there remains a degree of freedom in the solution corresponding to the variable $m_{16}$. This degree of freedom that remains in the solution was already noted by Han et al.\ [7]. The solution in Eq.\ (40) gives the prediction $\Delta_{\text{LHV}}=2(2p_8+2p_9+2p_{15}-1)$, so the Bell-CHSH inequality will be violated whenever $p_8+p_9+p_{15} >1$. Incidentally, we can easily prove the negativity of either $m_4$ or $m_{13}$ in this case by simply noting that $m_4+m_{13}= 1-p_8-p_9-p_{15}$.

In summary, in this paper we have proved the necessity of negative probability measures for the considered LHV model in all instances where the predictions by such a model are to give a violation of the Bell-CHSH inequality. Moreover, we have obtained the most general solution for LHV probability measures that reproduce any conceivable set of probabilities fulfilling the normalization condition and the causal communication constraint. We have observed that there remain seven degrees of freedom in the solution. In this respect, it should be emphasized that, as we have seen, it is only by imposing the condition of perfect correlation {\em and\/} fixing the eight corresponding LHV probability measures to zero, that the number of degrees of freedom remaining in the solution reduces to one. In general, however, the solution contains more than one degree of freedom. We remark that the achieved general solution for LHV probability measures can be used, specifically, to reproduce whichever quantum mecanical predictions for the probabilities $p_1,p_2,\ldots,p_{16}$ since, as was mentioned previously, the quantum theoretic predictions satisfy the consistency conditions of normalization and causal communication (see Ref.\ [13] and references therein).

Finally we note that, by taking a somewhat different approach, Rothman and Sudarshan [15] (see also Ref.\ [16]) have arrived at essentially the same results as those reached in this paper. Specifically, they show that the CHSH sum of correlations is derivable from a master probability distribution involving 16 joint probabilities for four ``simultaneous'' spin measurements along four axes (see Eq.\ (4.2) of [15]), and that the CHSH inequality is violated if the probabilities are allowed to become negative. They also give explicit solutions for the 16 four-probabilities, $P(++++)$, $P(+++-),\ldots$, $P(----)$ (see Tables II and III of [15]), that reproduce the standard quantum mechanical predictions.\footnote{%
For ease of comparison with the work of Rothman and Sudarshan in Ref.\ [15], here we write down the translation between their notation and ours for the 16 four-probabilities: $P(++++)\equiv m_1$, $P(+++-)\equiv m_2$, $P(+-++)\equiv m_3$, $P(+-+-)\equiv m_4$, $P(++-+)\equiv m_5$, $P(++--)\equiv m_6$, $P(+--+)\equiv m_7$, $P(+---)\equiv m_8$, $P(-+++)\equiv m_9$, $P(-++-)\equiv m_{10}$, $P(--++)\equiv m_{11}$, $P(--+-)\equiv m_{12}$, $P(-+-+)\equiv m_{13}$, $P(-+--)\equiv m_{14}$, $P(---+)\equiv m_{15}$, and $P(----)\equiv m_{16}$. From Table III of [15] we can check that, for a polarizer angle $\theta$ of $45^{\circ }$ (where $\theta=45^{\circ }$ is just the angle giving $\Delta=2\sqrt{2}$), we have $P(+-+-)=P(++-+)=P(++--)=P(+---)=P(-+++)=P(--++)=P(--+-)=P(-+-+)=(1-\sqrt{2})/16$ and $P(++++)=P(+++-)=P(+-++)=P(+--+)=P(-++-)=P(-+--)=P(---+)=P(----)=(1+\sqrt{2})/16$, in accordance with Eqs.\ (25) and (26).}

As cleverly anticipated by Feynman [17,18] (see also Ref.\ [11]), the important point arising from the discussions at hand can be summarized by saying that [15], ``The {\em only\/} difference between the classical and quantum cases is that in the former we assume the probabilities are positive-definite.'' Be that as it were, I would like to end this paper by quoting the final sentence ending the review paper in Ref.\ [3]: ``{\em Kolmogorov's axiom may hold or not}; {\em the probability for the existence of negative probabilities is not negative}'' (italics in the original).

\vspace{.2cm}
\textbf{Acknowledgements} --- I am grateful to Dimiter G. Chakalov for bringing my attention to the paper in Ref.\ [15].

\end{document}